\documentclass[11pt,a4paper]{article}
\usepackage{jheppub}
\usepackage{feynmp}
\usepackage{graphicx}
\title{Rapid field excursions and the inflationary tensor spectrum}
\author{Daniel Carney,}
\author{Willy Fischler,}
\author{Ely D. Kovetz,}
\author{Dustin Lorshbough,}
\author{Sonia Paban}

\affiliation{Theory Group, Department of Physics, University of Texas at Austin and \\ Texas Cosmology Center, \\ Austin, Texas 78712}

\emailAdd{carney@physics.utexas.edu}
\emailAdd{fischler@physics.utexas.edu}
\emailAdd{elykovetz@gmail.com}
\emailAdd{lorsh@utexas.edu}
\emailAdd{paban@physics.utexas.edu}

\abstract{We consider the effects of fields with suddenly changing mass on the inflationary power spectra. In this context, when a field $\chi$ becomes light, it will be excited. This process contributes to the tensor power spectrum. We compute these effects in a gauge-invariant manner, where we use a novel analytical method for evaluating the corrections to the tensor spectrum due to $\chi$ excitations. In the case of a scalar field, we show that the net impact on the tensors is small as long as the perturbative expansion is valid. Thus, in these scenarios, measurement of tensor modes is still in one-to-one correspondence with the Hubble scale.}


\arxivnumber{xxxx.xxxx}

\begin{document}
\unitlength = 1mm
\maketitle

\section{Introduction}
The energy scale of cosmic inflation is largely unknown. We know that it must be higher than about an MeV from nucleosynethesis and less than the Planck energy scale. Various assumptions can tighten this range, but experimental handles are clearly needed.

Currently, the most direct handle on the inflationary energy scale comes from (non)-detection of tensor fluctuations. This is because in typical models of inflation, the amplitude of tensor fluctuations is in one-to-one correspondence with the inflationary Hubble parameter, whereas scalar fluctuations are degenerate in $H$ and the slow-roll parameter $\epsilon$.\footnote{This is typically parametrized in terms of the ratio $r$ of tensor fluctuations to scalar fluctuations, the latter of which have been observed. This ratio is related to the scale of inflation by $r \sim 100 \times V_{\text{inflation}}/(10^{16} GeV)^4$; to date the non-detection of tensor fluctuations gives an upper bound $r \lesssim 0.2$, yielding $V_{\text{inflation}}^{1/4} \lesssim 10^{-2} M_{pl}$. \cite{Komatsu:2010fb}}

However, this conclusion has recently been challenged \cite{Cook:2011hg,Senatore:2011sp,Barnaby:2011qe,Barnaby:2012xt}. These authors have pointed out that although the amplitude of \emph{vacuum} tensor fluctuations uniquely probes the Hubble scale, it is conceivable that dynamics involving the metric could produce large correlations. Concretely, these authors have proposed that a spectator $\chi$ field (or fields) rapidly excited out of vacuum during inflation could produce significant radiative corrections to the tensor power spectrum.

The purpose of this paper is to clarify precisely in what way such dynamics can affect the tensor spectrum. Previous work has consisted of order-of-magnitude estimates \cite{Senatore:2011sp} or computations based on a perturbative expansion that differs from ours \cite{Cook:2011hg}; we elaborate on these differences in the text. Here, we show how to organize both the production of $\chi$ excitations and its effect on the tensor power spectrum in a systematic field theory computation. We develop an analytical method to compute the various perturbative contributions.

We find that any effects from rapidly excited scalar fields on the tensor power spectrum is small, as long as perturbation theory holds. The reason is that in order for the field to become excited, its mass must change rapidly on a timescale $\Delta t \ll H^{-1}$. Consistency of perturbation theory puts a lower bound on this timescale, restricting the non-adiabatic nature of the process. This leads to an upper bound on corrections to the tensor spectrum, rendering the corrections unobservable.

The rest of the paper is organized as follows. In section two, we review the inflationary background dynamics and production of $\chi$. In section three, we introduce our formalism for computing the tensor spectrum and state the results; the detailed computation is relegated to the appendix. In section four we make some comments regarding an alternative scenario in which $\chi$ has a nontrivial classical background, again concluding that significant effects on the tensors are generally ruled out. In section five we conclude.

\section{Background equations and $\chi$ production}
Consider gravity minimally coupled to two scalar fields, $\phi$ and $\chi$.  The action is
\begin{equation}
S = \frac{1}{2} \int d^4x \sqrt{-g} \left[ M^2 R + \partial_{\mu} \phi \partial^{\mu} \phi + \partial_{\mu} \chi \partial^{\mu} \chi - 2 V(\phi,\chi) \right].
\end{equation}
We assume that the potential is such that the inflaton $\phi$ has a large homogeneous part $\phi(x,t) = \overline{\varphi}(t) + \delta \phi(x,t)$ which sources a period of accelerated expansion, while $\chi$ acts only as a perturbation. The gravitational background sourced by $\overline{\varphi}$ is $\overline{g}_{00}(t) = 1, \overline{g}_{ij}(t) = -a(t)^2 \delta_{ij}$, i.e. we assume a flat FRW metric. As usual, $a(t)$ satisfies the Friedmann equation $H^2 M^2 = V(\overline{\varphi})/3$ and we take the de Sitter approximation $a(t) \approx e^{H t}$; here and throughout $M$ denotes the Planck mass.

We want to study the effects of excited quantum fields $\chi$ on the power spectrum of tensor perturbations. Inflation redshifts any such excitations away, so the initial conditions for field perturbations are the closest thing to a vacuum state that may be defined, the ``adiabatic vacuum,'' made precise shortly. Each mode
\begin{equation}
\hat{\chi}(\mathbf{x},t) = \int \frac{d^3k}{(2\pi)^3} e^{i\mathbf{k} \cdot \mathbf{x}} \hat{\chi}_{\mathbf{k}}(t), \ \ \ \ \hat{\chi}_{\mathbf{k}}(t) = \chi_{\mathbf{k}}(t) \hat{a}_{\mathbf{k}} + \overline{\chi_{-\mathbf{k}}(t)} \hat{a}_{-\mathbf{k}}^{\dagger}.
\end{equation}
is a damped harmonic oscillator with time-dependent frequency: 
\begin{equation}
\label{mode-eom}
\ddot{\chi}_k + 3 H \dot{\chi}_k + \tilde{\omega}^2 \chi_k = 0, \ \ \ \ \ \ \tilde{\omega}^2(k,t) = \left( \frac{k}{a(t)} \right)^2 + m^2(t).
\end{equation}
Here we used rotational invariance to write the modes in terms of $k = | {\mathbf{k}}|$. This equation can be mapped into an undamped oscillator by taking $\chi \mapsto a^{3/2} \chi$, except that this oscillator will have frequency
\begin{equation}
\label{frequency}
\omega^2 = \tilde{\omega}^2 - \frac{9}{4}H^2.
\end{equation}
The idea proposed in \cite{Senatore:2011sp,Cook:2011hg} is to give $\chi$ a time-dependent mass such that it becomes massless at some point during inflation. At low $k$ the mode's frequency will change non-adiabatically $| \dot{\omega}/\omega^2 | \gg 1$ and the mode will be excited. We assume that $m(t_0)=0$ at some time $t_0$ and take this to be a local minimum so that $m^2(t) \approx \dot{m}_0^2 (t-t_0)^2 = (k_{*}/a_0)^4 (t-t_0)^2$ near $t_0$. Here $a_0 = a(t_0)$ is the scale factor, and we have defined the (comoving) momentum scale $k_*$ which we will use to phrase all of our arguments.\footnote{The prototypical model \cite{Cook:2011hg} achieves this up by a coupling $g^2 \chi^2 (\phi-\phi_0)^2$ between $\chi$ and the inflaton. In these models, $k_* = a_0 \sqrt{g \dot{\overline{\varphi}}_0}$ where $\dot{\overline{\varphi}}_0$ denotes the background inflaton velocity at $t=t_0$. This model suffers from large corrections to the scalar power spectrum, so here we just leave the mass of $\chi$ as an unspecified function of time without positing its origin as a VEV of some other field.}

\begin{figure}[h]

\begin{center}$
\begin{array}{cc}

\includegraphics[scale=0.8]{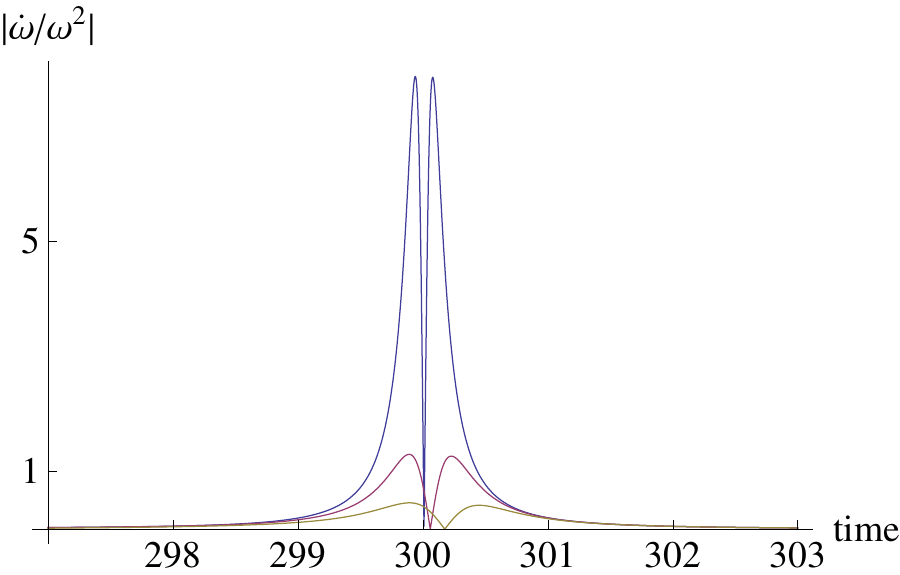}

&

\includegraphics[scale=0.8]{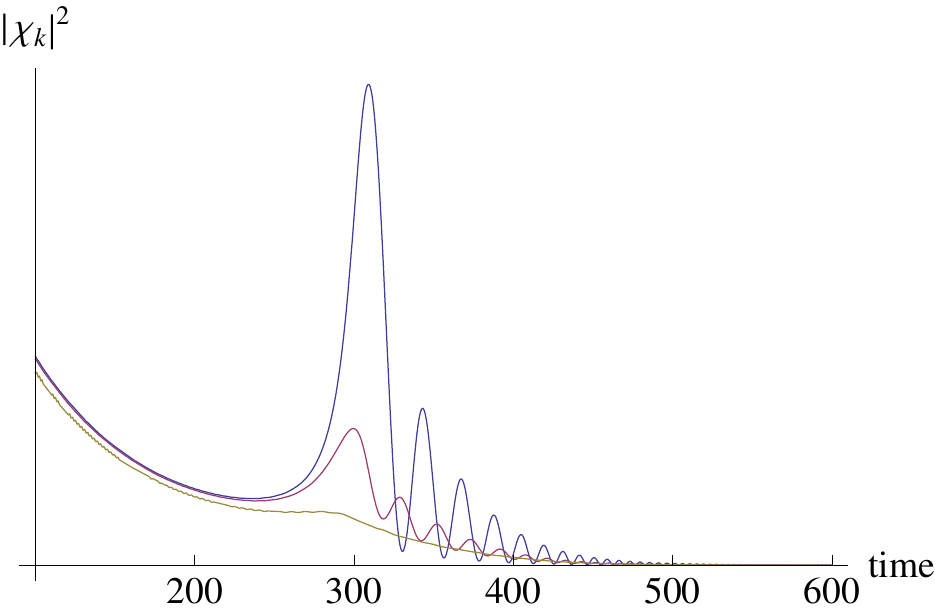}

\end{array}$
\end{center}
\caption{Left: Non-adiabaticity parameter $| \dot{\omega}/\omega^2|$ as a function of time for various wavelengths. Right: Corresponding numerical time evolution of the $\chi$ correlation function $\langle \hat{\chi}_k(t) \hat{\chi}_{-k}(t) \rangle = | \chi_k(t) |^2$. Long wavelengths (blue) have $| \dot{\omega}/\omega^2 | \gg 1$ near $t_0$ (in this picture, $t_0 \approx 300$); these modes undergo a sharp non-adiabatic evolution, leading to interference between the positive and negative frequency mode functions, cf. (\ref{chilate}). Shorter-wavelength modes have weaker effects (red, $k \approx k_{*}$) or evolve completely adiabatically (yellow, $k \gg k_{*}$). Notice that \emph{all} mode functions have a nontrivial evolution beyond the usual redshifting: they scale like $\omega^{-1/2}$, which is enhanced near $t_0$.}

\label{modes}
\end{figure}

For any comoving mode $k$, the time-variation of $\omega$ is very slow away from $t_0$ (fig. \ref{modes}). Thus the adiabatic mode function
\begin{equation}
\label{chi0}
\chi_k^0(t) = \frac{1}{\sqrt{2 (2\pi)^3 a^3(t) \omega(k,t)}} \exp{ \left\{ -i \int^t dt' \omega(k,t') \right\} }
\end{equation}
and its complex conjugate provide, to very good approximation, a basis of solutions to (\ref{mode-eom}), normalized to satisfy the Wronskian condition, see eg. \cite{Weinberg:2005vy}. Before the production event, $t \ll t_0$, we need to define the state of the $\chi$ field. High momentum modes with $k/a \gg m$ should match the Bunch-Davies behavior $\chi \sim 1/k$. Non-relativistic modes, $k/a \ll m$, will see the flat space vacuum, $\chi \sim 1/m$. The adiabatic mode function (\ref{chi0}) reduces to both of these limits. According to the usual argument (the equivalence principle) we therefore take as an initial condition $\chi_k = \chi_k^0$ and set the state to be the one defined by $a_k \mid 0 \rangle_{\chi} = 0$. 

From eq. (\ref{frequency}), we see that the behavior of the modes of $\chi$ near $t=t_0$ depends on the wavelength. Very high momentum modes have real and slowly-varying frequency for all times and thus evolve adiabatically. The more interesting case is that of modes with long wavelengths, $k/a_0 H \ll \infty$, so that the frequency evolves rapidly $|\dot{\omega}/\omega^2| \gg 1$. These modes evolve non-adiabatically near $t_0$, (fig. \ref{modes}). The timescale of this process is $\Delta t = a_0 / k_{*}$; we assume that this is shorter than the Hubble timescale, i.e. $\Delta t \ll H^{-1}$.

Sufficiently long after $t_0$, every mode again evolves adiabatically and so the general solution to (\ref{mode-eom}) must be a superposition of the adiabatic solution and its conjugate,
\begin{equation}
\label{chilate}
\chi_k(t) = \alpha(k,t) \chi_k^0(t) + \beta(k,t) \overline{\chi_k^0(t)}.
\end{equation}
The coefficients $\alpha$ and $\beta$ are the usual Bogoliubov coefficients; the Wronskian condition reduces to $|\alpha|^2 - |\beta|^2 = 1$. The initial condition described above corresponds to $\alpha = 1, \beta = 0$ as $t \rightarrow -\infty$. Modes that evolve non-adiabatically pick up a non-zero $\beta$ around $t=t_0$ and then maintain this value for $t > t_0$. 

A late-time observer, viewing physics with respect to her naturally defined adiabatic vacuum,\footnote{i.e. the state defined by $b_k \mid \tilde{0} \rangle = 0$ where $b_k$ is the Bogoliubov-transformed annihilation operator. \cite{Birrell:1982ix}} would see the $\chi$ field modes excited with occupation numbers $n_k = | \beta(k) |^2$. In order to quantify the excitations of $\chi$, we thus want to find the value of $\beta$ at late times. The simplest way is to solve the mode equation numerically, fig. \ref{modes}. One can also find approximate analytic solutions and match these to (\ref{chilate}), see eg. \cite{Cook:2011hg,Kofman:1997yn}. 

Another approach, developed in \cite{Chung:1998bt}, is to consider the analytical continuation of $\omega$ to complex times. Later, we will generalize this approach to compute our loop diagrams. It can be shown that, if $| \dot{\omega}/\omega^2 | \ll 1$ for all time, then at any time $t$ one can compute $\beta$ by the integral \cite{Chung:1998bt}
\begin{equation}
\label{betaformal}
\beta(k,t) = \int_{-\infty}^t dt' \frac{\dot{\omega}(k,t')}{2 \omega(k,t')} \exp{ \left\{ -2i \int_{-\infty}^{t'} dt'' \omega(k,t'') \right\}}.
\end{equation}
Note that for $| t - t_0 | \gg \Delta t$, $\dot{\omega}/\omega$ is small and the large $\omega$ makes the phase incoherent, so $\beta$ is constant. The point is that near $t_0$ this integral would naively pick up a non-trivial contribution; the problem is that the integral does not make sense in this region because the adiabaticity condition $| \dot{\omega}/\omega^2 | \ll 1$ is violated. The solution is to adapt a trick from 1D Schrodinger scattering problems: we can deform the contour of integration here off of the real axis in such a way that $| \dot{\omega}/\omega^2 | \ll 1$ along the entire contour. The frequency $\omega$ is a square root and thus has a branch cut on the complex plane, and as we go around this cut, the frequency picks up an imaginary part, so that the exponential picks up a real part. We give a detailed account of this method in the appendix. For the time being, the main point is that both the matching and analytical continuation methods have been used to find $\beta(t \rightarrow +\infty)$, and they agree, yielding
\begin{equation}
\label{beta}
\beta(t \rightarrow \infty) = \frac{\pi}{3} \exp \left\{ -\frac{\pi}{2} \left[ \left( \frac{k}{k_*} \right)^2 - \left( \frac{3}{2} \frac{a_0 H}{k_*} \right)^2  \right] \right\}.
\end{equation}
This formula is the key to our result that the effect on the tensor spectrum is small. The tensors are derivatively coupled to $\chi$ but $\chi$ is produced in the infrared, so loop effects are localized to a small and infrared portion of phase space.\footnote{Previous studies neglected the expansion of the universe by simply setting $H=0$. For subhorizon modes $k \gg a_0 H$ this approximation is good as can be seen easily from the above formula since by assumption of non-adiabaticity $\Delta t^{-1} = k_* \gg H$; the derivative couplings of $\chi$ to the gravitons mean that this is also a good approximation for our problem.}

\section{Computation of the tensor spectrum}

The basic observable quantity we are interested in is the two-point correlation function of the tensor fluctuations, defined by $g_{\mu\nu}(x,t) = \overline{g}_{\mu\nu}(t) + h_{\mu\nu}(x,t)$. Here $\overline{g}$ is the background inflationary metric described earlier. One typically defines the tensor power spectrum as
\begin{equation}
\Delta_t^2(k) = \frac{k^3}{2\pi^2} P_t(k)
\end{equation}
where $P_t(k)$ is the two-point correlation function of the tensor fluctuations,
\footnote{The Fourier conventions here are
\begin{equation}
\hat{h}_{ab}(x,t) = \sum_{s = \pm} \int \frac{d^3k}{(2\pi)^3} e^{ikx} \epsilon_{ab}^s(k) \hat{h}^s_k(t), \ \ \ \ \hat{h}^s_k(t) = h^s_{k}(t) \hat{b}^s_k + \overline{h_{-k}^s(t)} \hat{b}_{-k}^{s \dagger}.
\end{equation}
We will work in the transverse-traceless gauge for the gravitons. The polarization tensors satisfy\cite{Weinberg:2008zzc}
\begin{eqnarray*}
\sum_s \epsilon_{ab}^s(k) \epsilon_{cd}^s(-k) & = & \delta_{ac} \delta_{bd} + \delta_{ad} \delta_{bc} - \delta_{ab} \delta_{cd} + \delta_{ab} \hat{k}_c \hat{k}_d + \delta_{cd} \hat{k}_a \hat{k}_b \\
& & - \delta_{ac} \hat{k}_b \hat{k}_d - \delta_{ad} \hat{k}_b \hat{k}_c - \delta_{bc} \hat{k}_a \hat{k}_d - \delta_{bd} \hat{k}_a \hat{k}_c + \hat{k}_a \hat{k}_b \hat{k}_c \hat{k}_d.
\end{eqnarray*}}
\begin{equation}
(2\pi)^3 P_t(k) \delta(\mathbf{k}+\mathbf{k'}) = \sum_{s} \langle \hat{h}_\mathbf{k}^s \hat{h}_{\mathbf{k'}}^{s} \rangle (t).
\end{equation}
This is an in-in computation which we perform largely following \cite{Weinberg:2005vy}. We work in the interaction picture, so that all field operators evolve according to their free\footnote{Here we are treating the $\chi$ field as having a time-dependent mass; of course if this mass comes from eg. coupling $\chi$ to $\overline{\varphi}$ we still include this coupling in the ``free'' equations of motion.} equations of motion while the state evolves according to the time evolution determined by the interaction Hamiltonian. In other words, for any operator $\mathcal{O}$, we have
\begin{equation}
\label{inin1}
\langle \Psi \mid \mathcal{O} \mid \Psi \rangle (t) = \langle \Psi \mid U^{-1}(t_i,t) \mathcal{O}(t) U(t_i,t) \mid \Psi \rangle,
\end{equation}
where the explicit time-dependence of the operator is that of the interaction picture, and
\begin{equation}
\label{inin2}
U(t_i,t) = T \exp \left\{-i \int_{t_i}^t dt'\ H_{int}(t') \right\}
\end{equation}
where here $H_{int}$ has both explicit time-dependence as well as time-dependence from writing all field operators in the interaction picture. The state $\mid \Psi \rangle$ is determined at the initial time $t_i$. In practice we are going to take $t_i \rightarrow -\infty$ and set the state to be $\mid 0 \rangle$, the state annihilated by the annihilation operators of all the relevant fields, as described above. 

In our problem the relevant part of the interaction Hamiltonian comes from expanding the Lagrangian for $\chi$ minimally coupled to gravity, giving the interactions (see eg. \cite{Park:2011ww})
\begin{equation}
\mathcal{L}_{int} = \sqrt{-\overline{g}} \left[\frac{1}{2} h^{\mu \nu} \partial_{\mu} \chi \partial_{\nu} \chi + \frac{1}{8} h^{\mu \nu} h_{\mu \nu} \overline{g}^{\rho \sigma} \partial_{\rho} \chi \partial_{\sigma} \chi - \frac{1}{2} h^{\mu \rho} h_{\rho}^{\nu} \partial_{\mu} \chi \partial_{\nu} \chi \right].
\end{equation}
Here we have dropped the terms that vanish for transverse, traceless gravitons. The interaction picture fields have time-evolution governed by their free equations of motion. Ignoring the Planck-suppressed effects of $\chi$ on the tensors, each polarization is a massless minimally coupled scalar field in de Sitter,
\begin{equation}
h_k^s(t) = \frac{H}{M \sqrt{2 k^3}} \left(1 + i \frac{k}{aH}\right) e^{-i \frac{k}{aH}}.
\end{equation}
Meanwhile $\chi$ has a more complicated evolution due to its interaction with the background inflaton field, as described above. The object of interest is $\langle \Psi | \hat{h}_k^s \hat{h}_{k'}^{s} | \Psi \rangle (t)$. From here out we drop the explicit reference to the state to maintain some sanity in the notation. Using (\ref{inin1}), (\ref{inin2}) and Wick's theorem,\footnote{Wick's theorem for time-ordered vacuum correlators is of course $\langle T \prod_i \hat{\mathcal{O}}_i(t_i) \rangle = \sum_{ij} \langle T \hat{\mathcal{O}}_i(t_i) \hat{\mathcal{O}}_j(t_j) \rangle$.} we can expand this in a kind of Dyson series, in powers of the interaction Hamiltonian. Following the logic in \cite{Weinberg:2005vy}, it is possible to develop a Feynman diagram-style calculus to organize this computation. Unlike Weinberg we work in Fourier space directly, but otherwise the procedure is identical. 

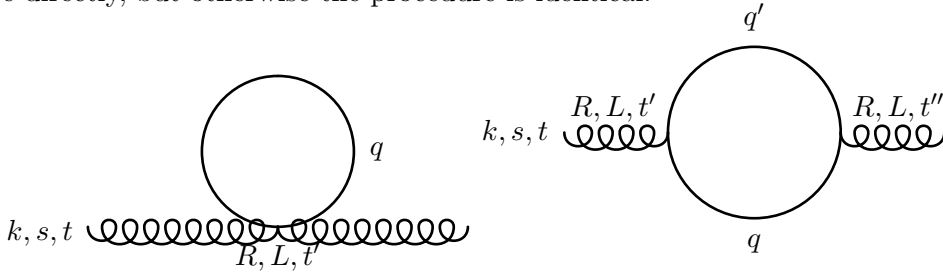
\begin{figure}[h]

\begin{center}$
\begin{array}{cc}

\begin{fmffile}{diag1}
\fmfframe(5,0)(5,0){
\begin{fmfgraph*}(50,20)
\fmfstraight
\fmfbottom{a,b,c}
\fmftop{d,e,f}
\fmf{gluon}{a,b}
\fmf{gluon}{b,c}
\fmf{plain,tension=.5,label=$q$,right}{b,e}
\fmf{plain,tension=.5,right}{e,b}
\fmflabel{$R,L,t'$}{b}
\fmflabel{$k,s,t$}{a}
\end{fmfgraph*}
}
\end{fmffile}

&

\begin{fmffile}{diag2}
\fmfframe(5,0)(5,0){
\begin{fmfgraph*}(50,25)
\fmfstraight
\fmfleft{a}
\fmfright{b}
\fmflabel{$k,s,t$}{a}
\fmf{plain,tension=.3,label=$q$,right}{v1,v2}
\fmf{plain,tension=.3,label=$q'$,right}{v2,v1}
\fmf{gluon}{a,v1}
\fmf{gluon}{v2,b}
\fmfv{label=$R,,L,,t'$,label.angle=135}{v1}
\fmfv{label=$R,,L,,t''$,label.angle=45}{v2}
\end{fmfgraph*}
}
\end{fmffile}	

\end{array}$
\end{center}
\caption{Diagrams contributing to $\langle h_k h_k \rangle$ at leading order ($1/M^2$). The scalar lines are $\chi$ correlators and the curly lines are gravitons. Later we denote the first diagram by $A$ and the second by $B$.}

\label{diagrams}

\end{figure}

The standard result for the tensor spectrum comes from assuming that the metric fluctuations are in the Bunch-Davies state defined above and ignores interactions. In the language we are using here, this means it is just the tree-level correlation function:
\begin{equation}
\langle \hat{h}_{\mathbf{k}} \hat{h}_{\mathbf{k'}} \rangle(t) = 
	\begin{fmffile}{diag3}
	\begin{fmfgraph}(30,5)
	\fmfleft{a}
	\fmfright{b}
	\fmf{gluon}{a,b}
	\end{fmfgraph}
	\end{fmffile}
= \delta(\mathbf{k}+\mathbf{k'}) | h_k(t)^2 | = \frac{H^2}{2 M^2 k^3} \left(1+ \left( \frac{k}{a H} \right)^2 \right).
\end{equation}
At late times $t \rightarrow \infty$ this gives the usual scale-invariant spectrum of inflation. Corrections to this come from insertions of the interaction Hamiltonian in (\ref{inin1}). In our problem this means that the first non-vanishing corrections will come at order $1/M^2$.

There are two diagram topologies contributing at first order, see fig. \ref{diagrams}. Adapting the rules from  \cite{Weinberg:2005vy} or just brute-force expanding (\ref{inin1}), these can be evaluated in terms of integrals over time and loop momenta. Each vertex is labelled by a time argument which is integrated over from $t \rightarrow -\infty$ until the time of measurement $t$ (in practice we measure correlators well outside the horizon, $t \rightarrow \infty$; we have checked numerically that the loop corrections become very nearly constant after horizon exit of the ``external'' graviton momentum $k$, as with standard loop corrections in inflation). For each vertex, we may have an insertion on either the right or the left in (\ref{inin1}), and we must sum over all of these possibilities. The loops of $\chi$ fluctuations contain the information about $\chi$ excitations because the propagators are enhanced as explained in the previous section.

The first diagram, call it $A$, is a tadpole and is straightforward to evaluate. This diagram has been missed by previous treatments of the problem. Summing over polarizations $s$ and adding the diagram with a right insertion to the one with a left insertion, one obtains
\begin{equation}
\label{tadpole}
\sum_{s,R,L} \langle \hat{h}_\mathbf{k}^s \hat{h}_{\mathbf{k'}}^{s} \rangle = - i \delta(\mathbf{k}+\mathbf{k'}) \int_{-\infty}^t \frac{dt' d^3q}{(2\pi)^3} f_4(q,t') \text{Im} \left( h_k^s(t) \overline{h_k^s(t')} \right)^2 | \chi_q(t') |^2
\end{equation}
where $f_4$ comes from the sum over polarizations. The second diagram $B$ has four combinations of right/left choices. Summing over polarizations again, we find for the right-right diagram
\begin{eqnarray}
\label{diag2-1}
\lefteqn{ \sum_s \langle \hat{h}_\mathbf{k}^s \hat{h}_{\mathbf{k'}}^{s} \rangle_{RR} = - \delta(\mathbf{k}+\mathbf{k'}) \int_{-\infty}^t dt' \int_{-\infty}^{t'} dt'' \int \frac{d^3q}{(2\pi)^3} } \nonumber \\
& &\\
& & \times f_3(q,t',t'') h_k^s(t) \overline{h_k^s(t')} h_k^s(t) \overline{h_k^s(t'')} \chi_q(t') \overline{\chi_q(t'')} \chi_{q'}(t') \overline{\chi_{q'}(t'')}  \nonumber
\end{eqnarray}
where $\mathbf{k} + \mathbf{q} + \mathbf{q'} = 0$ and $f_3$ is again from the polarization sum. The left-left diagram is just the conjugate of this, $\langle \hat{h}_\mathbf{k}^s \hat{h}_{\mathbf{k'}}^{s} \rangle_{LL} = \overline{\langle \hat{h}_\mathbf{k}^s \hat{h}_{\mathbf{k'}}^{s} \rangle_{RR}}$. The mixed-vertex diagrams contribute equally,
\begin{eqnarray}
\label{diag2-2}
\lefteqn{ \sum_s \langle \hat{h}_\mathbf{k}^s \hat{h}_{\mathbf{k'}}^{s} \rangle_{RL} = +\delta(\mathbf{k}+\mathbf{k'}) \int_{-\infty}^t dt' \int_{-\infty}^{t'} dt'' \int \frac{d^3q}{(2\pi)^3} } \nonumber \\
& & \\
& & \times  f_3(q,t',t'') h_k^s(t') \overline{h_k^s(t)} h_k^s(t) \overline{h_k^s(t'')} \chi_q(t') \overline{\chi_q(t'')} \chi_{q'}(t') \overline{\chi_{q'}(t'')}, \nonumber
\end{eqnarray}
and $\langle \hat{h}_k^s \hat{h}_{k'}^{s} \rangle_{LR} = \langle \hat{h}_k^s \hat{h}_{k'}^{s} \rangle_{RL}$. Explicitly, the functions $f_3$ and $f_4$ are given by
\begin{equation}
\label{f3}
f_3(q,t',t'') = a(t') a(t'') \left[q^4 - 2 q^2 (\mathbf{q} \cdot \hat{\mathbf{k}})^2 + (\mathbf{q} \cdot \hat{\mathbf{k}})^4 \right],
\end{equation}
\begin{equation}
\label{f4}
f_4(q,t') = f_4^a(q,t') + f_4^b(q,t'),
\end{equation}
\begin{equation}
f_4^a = a(t') \left[ 5 q^2 - 2(\mathbf{q} \cdot \hat{\mathbf{k}})^2 \right], \ \ f_4^b = \frac{1}{a(t')} \left[ \left( \frac{3H}{2} + \frac{\dot{\omega}}{2 \omega} \right)^2 + \omega^2 \right].
\end{equation}
where the inner products are with respect to $\delta_{ab}$.

The ultraviolet divergences of these loop integrals need to be regularized. Previous treatments have done this by a process called ``adiabatic subtraction'' in which one makes replacements like $|\chi_q(t)|^2 \rightarrow |\chi_q(t)|^2 - |\chi_q^0(t)|^2$. For sufficiently high $q$, this difference vanishes since the modes are evolving adiabatically, and so all loops become ultraviolet-finite. However, this procedure has the disadvantage of affecting answers in the infrared regime we are interested in \cite{Durrer:2009ii,Agullo:2009vq}. Moreover, the intuition is that we are ``subtracting off the adiabatic behavior,'' a notion which seems difficult in our problem since the infrared modes are in no sense evolving adiabatically (i.e. $\chi^0_q$ is not even approximately a solution of the equations of motion).

In the appendix we present a simple method for computing the finite effects of the loop. If we analytically continue the time arguments to the complex plane, we can take the time integrals to run over contours such that along the entire contour, $|\dot{\omega}/\omega^2| \ll 1$ for all momenta. This means that we can use the adiabatic mode functions (\ref{chilate}) for the entire time integral. This simplifies the computations and allows us to track which parts of the loops come from $\chi$ excitations, namely those terms proportional to $\beta$ (recall that a mode that is not excited has $\beta = 0$ for all time). Focusing on these terms also regularizes the momentum integrals because $\beta \sim e^{-q^2/k_*^2}$.\footnote{This process is actually equivalent to adiabatic subtraction for diagram $A$, but we note that adiabatic subtraction is still only sensible once we have deformed the integration contour so that we actually have adiabatic solutions for all time.}

\begin{figure}[h]
\label{FandG}
\begin{center}$
\begin{array}{cc}

\includegraphics[scale=0.8]{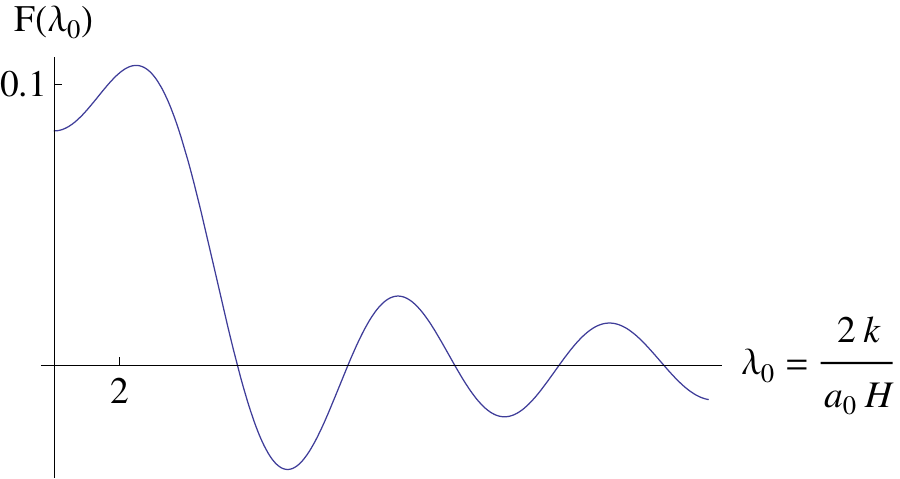}

&

\includegraphics[scale=0.8]{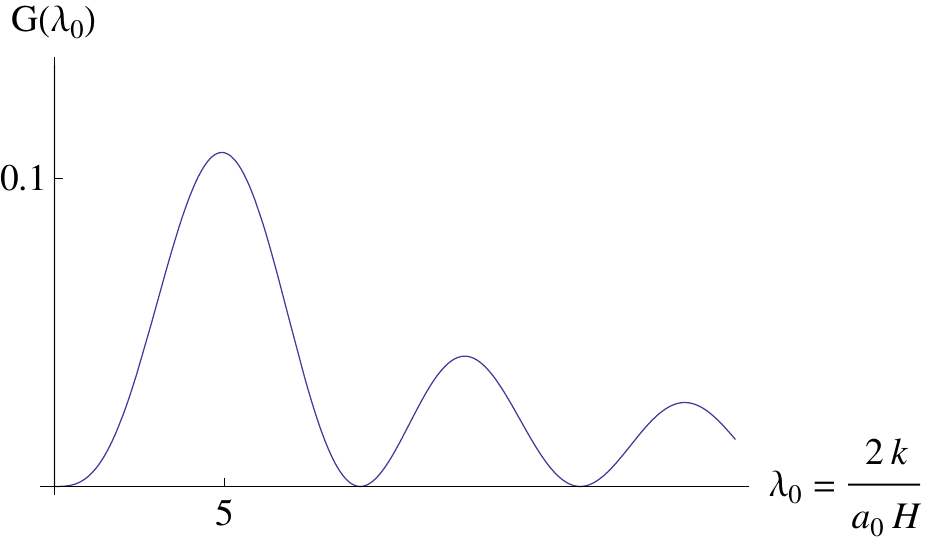}

\end{array}$
\end{center}
\caption{Dimensionless functions $F$ and $G$ appearing in the answer. $F$ enters into the tadpole diagram while $G$ enters into the other diagram.}
\end{figure}

The result of these computations is easy to state. The first diagram topology contributes at most
\begin{equation}
|A| \lesssim \mathcal{O}(1) \times \frac{H^2}{2 M^2 k^3} \frac{k_*^3}{H M^2} F \left( \frac{2k}{a_0 H} \right) ,
\end{equation}
while the second diagram topology gives
\begin{equation}
|B| \lesssim \mathcal{O}(1) \times \frac{H^2}{2 M^2 k^3} \frac{k_*^3}{H M^2} G \left( \frac{2k}{a_0 H} \right).
\end{equation}
The dimensionless functions $F$ and $G$ appearing here are given explicitly in the appendix, eq. (\ref{F}) and (\ref{G}); we plot them in fig. \ref{FandG} for convenience. The key is that they are both $\mathcal{O}(10^{-1})$ and peaked for $k \approx a_0 H$, i.e. the effect on the tensor spectrum is dominated for tensor modes whose wavelength is roughly the size of the horizon at the time $\chi$ becomes massless.

In the end, this means that the complete tensor power spectrum, computed to one loop and including only the effects from $\chi$ excitations, is given by (at most)\footnote{The term involving $G$ is precisely the one found in \cite{Cook:2011hg}.}
\begin{equation}
P_t(k) = \frac{H^2}{M^2} \left[1 + \frac{k_*^3}{H M^2} \left( F \left( \frac{2 k}{a_0 H} \right) + G\left( \frac{2 k}{a_0 H} \right) \right) \right].
\end{equation}
Let us estimate the size of this correction. The energy density contained in the $\chi$ excitations (normalizing $a=1$) should be roughly $E^4 = \int d^3q\ n_q \omega$ where $n_q = |\beta|^2 \sim e^{-q^2/k_*^2}$ is the occupation number for the $q$th mode. We have roughly $\omega \sim q$ and so this integral gives $E^4 \sim k_*^4$. In order for inflation to proceed, this energy should be subdominant to the inflaton potential, so we have $E^4 \sim k_*^4 \ll V_{inf} \sim H^2 M^2$ via the Friedmann equation, thus $k_*^2/M^2 \ll H/M$. This means that the correction to the power spectrum is bound from above by
\begin{equation}
\frac{k_*^3}{H M^2} \ll \frac{k_*}{M}.
\end{equation}
This quantity had better be quite a bit less than unity since otherwise we are exciting $\chi$ quanta with energy of order the Planck mass. Thus, this correction is quite small.

\section{Classical $\chi$ backgrounds}

\begin{figure}[h]
\label{diagrams2}
\begin{center}$
\begin{array}{cc}

\begin{fmffile}{diag2-1}
\fmfframe(5,0)(5,0){
\begin{fmfgraph*}(50,20)
\fmfstraight
\fmfbottom{a,b,c,d}
\fmftop{e,f,g,h}
\fmf{gluon}{a,b}
\fmf{dashes}{b,c}
\fmf{gluon}{c,d}
\fmf{dbl_dashes}{b,f}
\fmf{dbl_dashes}{c,g}
\fmfv{decoration.shape=cross}{f}
\fmfv{decoration.shape=cross}{g}
\end{fmfgraph*}
}
\end{fmffile}

&

\begin{fmffile}{diag2-2}
\fmfframe(5,0)(5,0){
\begin{fmfgraph*}(50,20)
\fmfstraight
\fmfbottom{a,b,c}
\fmftop{d,e,f,g}
\fmf{gluon}{a,b}
\fmf{gluon}{b,c}
\fmf{dbl_dashes}{b,e}
\fmf{dbl_dashes}{b,f}
\fmfv{decoration.shape=cross}{e}
\fmfv{decoration.shape=cross}{f}
\end{fmfgraph*}
}
\end{fmffile}

\\ \\ \\ \\

\begin{fmffile}{diag2-3}
\fmfframe(5,0)(5,0){
\begin{fmfgraph*}(50,20)
\fmfstraight
\fmfbottom{a,b,c,d}
\fmftop{e,f,g,h}
\fmf{plain}{a,b}
\fmf{dashes}{b,c}
\fmf{plain}{c,d}
\fmf{dbl_dashes}{b,f}
\fmf{dbl_dashes}{c,g}
\fmfv{decoration.shape=cross}{f}
\fmfv{decoration.shape=cross}{g}
\end{fmfgraph*}
}
\end{fmffile}

&

\begin{fmffile}{diag2-4}
\fmfframe(5,0)(5,0){
\begin{fmfgraph*}(50,20)
\fmfstraight
\fmfbottom{a,b,c}
\fmftop{d,e,f,g}
\fmf{plain}{a,b}
\fmf{plain}{b,c}
\fmf{dbl_dashes}{b,e}
\fmf{dbl_dashes}{b,f}
\fmfv{decoration.shape=cross}{e}
\fmfv{decoration.shape=cross}{f}
\end{fmfgraph*}
}
\end{fmffile}

\end{array}$
\end{center}
\caption{Some typical diagrams involving background $\chi$ fields, insertions of which we denote with crosses and double lines. Dashed lines are $\delta \chi$ propagators and solid lines are $\zeta$ propagators.}
\end{figure}
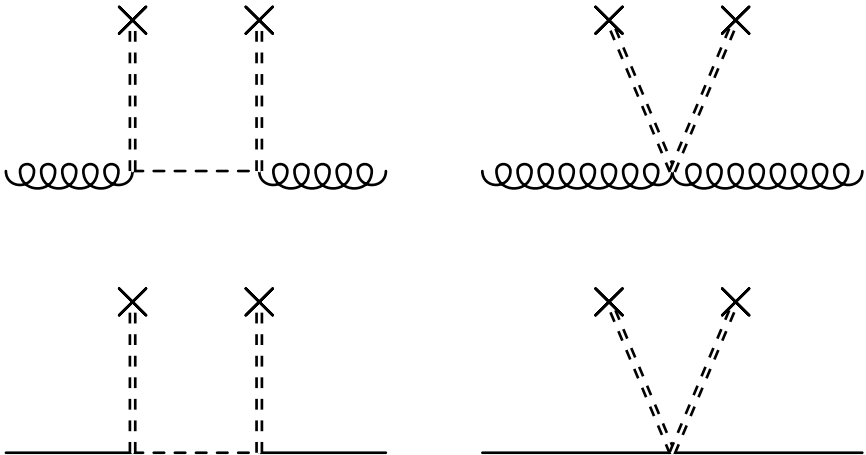

If the quantum fluctuations of a field $\chi$ become large enough they behave classically.  We now consider the case whereby quantum fluctuations have been amplified through some unspecified process enough to become classical. 

The slow roll constraints imply that the inflaton potential is much greater than the derivative terms of the $\chi$ field. The corrections to the tensor power from the inflaton are proportional to the inflaton potential whereas the corrections from the $\chi$ background are proportional to derivatives of $\chi$. The scalar power spectrum depends on derivatives of the $\chi$ background at the same order. Conversely, the scalar spectrum is more sensitive to the inflaton potential than the tensors. Thus, the corrections to the tensor-to-scalar ratio $r$ are more sensitive to the inflaton potential and are relatively insensitive to the background $\chi$ field. 

The corrections may overcome this suppression given an enormous number of spectator fields, however then one must question the use of perturbation theory for studying such a system since each field will contribute to the interaction Hamiltonian through at least a gravitational coupling. Therefore inference of the inflationary scale $H$ based on the positive detection of tensor modes is resilient to the effects of introducing additional  scalar fields.

\section{Conclusions}

Detection of tensor modes generated during inflation is currently our best hope of determining the energy scale of inflation, which to date is largely unknown. In the simplest models of inflation, in which tensor vacuum fluctuations dominate, a measurement of the tensors is in one-to-one correspondence with the energy scale. It is conceivable that dynamics in more complicated scenarios could destroy this correspondence.

Corrections to the tensor power spectrum will generally lead to corrections to the scalar power spectrum.  Recently it has been proposed that sudden excitation out of vacuum of some spectator $\chi$ fields could lead to enhancements of the tensor spectrum without significantly affecting the scalar spectrum.

In this work, we have given a systematic field theoretical treatment of both the excitation of $\chi$ and its effects on the tensor spectrum. Using this framework, we have studied the case in which $\chi$ is a scalar field. We studied the case when the field is excited at a specific time, where we have shown that there is a self-consistent upper bound on the corrections to the tensor power spectrum. This upper bound constrains the effect to be unobservable.

These results are consistent with previous studies. It has been suggested recently that modifications to this setup, especially the spin of the $\chi$ field, could change the answers we have obtained here. In our language, this is conceivable since it could change the powers of the momentum scale $k_*$ appearing in the correction to the tensor spectrum. In scenarios with certain vector couplings, gauge quanta can be produced continuously rather than suddenly \cite{Barnaby:2012xt}. One could consider the case in which there is a large number of $\chi$ fields, as in axion-based models of inflation \cite{Senatore:2011sp,Barnaby:2011qe}; our computation suggests that perturbation theory will break down with a large number of these fields, and new methods will have to be applied. However, we do conclude that having one (or even quite a few) spectator scalar $\chi$ fields suddenly excited during inflation will not produce a significant effect on the tensor power spectrum, and so the detection of tensor modes in these scenarios is still in one-to-one correspondence with the energy scale of inflation.

\acknowledgments

We thank Joel Meyers for useful discussions. This material is based upon work supported by the National Science Foundation under Grant No. PHY-0969020 and by the Texas Cosmology Center, which is supported by the College of Natural Sciences and the Department of Astronomy at the University of Texas at Austin and the McDonald Observatory.

\appendix
\section{Evaluation of loop integrals}

\subsection{General setup}

\begin{figure}
  \centering
    \includegraphics[scale=1]{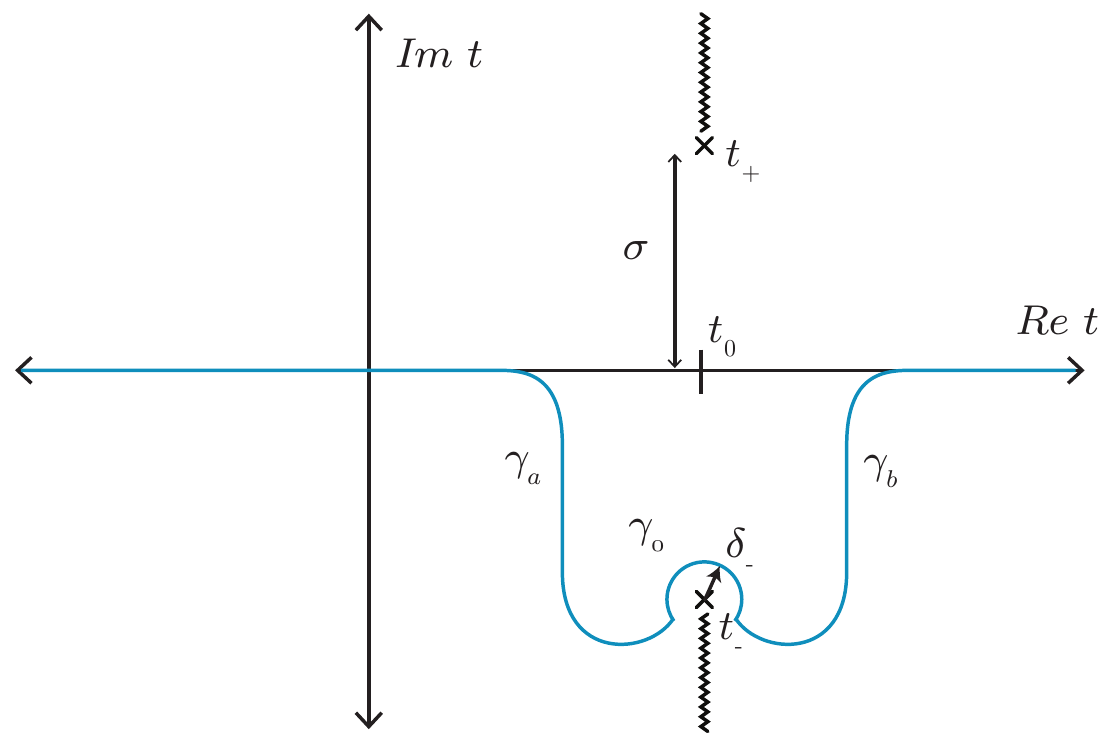}
  \caption{Contour used for evaluating $\beta$.}
  \label{contour}
\end{figure}

Because the $\chi$ field evolves non-adiabatically near $t_0$, it cannot always be expressed in terms of positive and negative frequency mode functions. In computing certain observables this is at best a technical headache, but in particular for loop integrals this is problematic because it makes it difficult to extract the part of the loop involving excited or ``particle'' states. 

In order to circumvent these issues, we can extend a trick developed in \cite{Chung:1998bt}. The key idea is that the non-adiabatic behavior of $\chi$ is characterized by the fact that $| \dot{\omega}/\omega^2 | \gg 1$, but only in a small region of time around $t_0$. If we analytically continue in $t$, then we can integrate along a contour $\gamma = \gamma(t)$ in the complex $t$-plane such that $| \dot{\omega}/\omega^2 | \ll 1$ for all $t$. Then the adiabatic mode expansion is valid for the entire integral.

As it turns out, once the integrals are analytically continued in this way, they can be evaluated by a steepest descent method. Generally, in evaluating $\beta$ using (\ref{betaformal}) or the time integrals in the loop diagrams, we encounter integrals of the form
\begin{equation}
\label{generaltimeint}
I = \int_{-\infty}^{+\infty} dt' \ f(\omega(t'),t') e^{-2 i \int_{-\infty}^{t'} \omega(t'')  dt'' }
\end{equation}
where $f$ contains some (negative) powers of $\omega$ and the time-dependence in $f$, other than the frequency terms, is much slower than the time-variation of $\omega$. Although $\omega$ is real and positive for all real $t$, its analytic continuation has two zeroes. Then the integrands contain branch cuts from the square root in $\omega$ as well as (sometimes) poles from factors of $\omega^{-1}$. We can thus take the contour $\gamma$ along which we perform this integral to be a steepest descent contour localized near these points because the exponential is oscillating rapidly except very close to the points where $\omega \approx 0$.

Let us set up the analytic continuation and steepest descent contours. Near $t_0$ and in the slow roll approximation we have
\begin{equation}
\label{beta2}
\omega \approx \sqrt{\frac{k^2}{a_0^2} + \frac{k_{*}^4}{a_0^4}(t-t_0)^2 - \left( \frac{3H}{2} \right)^2} = \frac{k_{*}^2}{a_0^2} \sqrt{(t-t_-)(t-t_+)}.
\end{equation}
Here we have assumed that that period of non-adiabaticity is shorter than a Hubble time so that $a(t) \approx a_0 = a(t_0)$, and as before $k_* = \Delta t^{-1} \gg a_0 H$ is the comoving momentum scale of the problem. We have factored $\omega$ by its zeroes 
\begin{equation}
\label{zeroes}
t_{\pm} = t_0 \pm \sigma, \ \ \ \ \ \sigma = \frac{a_0^2}{k_*^2} \sqrt{ \left( \frac{3H}{2} \right)^2 - \frac{k^2}{a_0^2}}.
\end{equation}
The branch cut(s) of this function must be taken with some care. As described in section 2, there are different types of behavior as a function of the momentum $k$. For very low momentum modes, $\sigma$ is real and $\omega$ has its zeroes on the real axis. However, for high momentum modes, $k/a_0 H > 3/2$, $\sigma$ is pure imaginary, and the zeroes of $\omega$ are symmetrically distributed across the real line; these are the modes that affect the tensor spectrum and we concentrate on them from here out. When we analytically continue $\omega$, we have to choose a branch cut to make this a well-defined function, such that $\omega$ is real and positive along the real axis. The contour of integration can be deformed into a steepest descent curve as long as we do not cross the branch cuts.\footnote{Many authors write that because the metric's expansion is assumed slow compared to the timescale of the excitations, it is consistent to directly set $H = 0$ in the frequency. This is not quite correct, but the phase space of the excited modes is enormously larger than the phase space outside the horizon, and indeed the excited modes dominate the loop integrals for exactly this reason.}

Consider integrals of the form (\ref{generaltimeint}), with $\omega$ continued as just described. Let us construct a good steepest descent contour $\gamma$ along which to evaluate the time integrals. First break up the contour as $\gamma = \gamma_a + \gamma_0 + \gamma_b$ where $\gamma_{a,b}$ are the contours leading into and out of the branch point, i.e. from $-\infty$ to $t_{\pm}$ and from $t_{\pm}$ to $+\infty$, respectively, and $\gamma_0$ runs from $t_{\pm}$ to $t$ near $t_{\pm}$. The exponential becomes a product and so
\begin{equation}
I = \int_{\gamma} dt' f(\omega(t'),t') e^{ -2i \int_{\gamma_a} \omega dt'' } e^{ -2i \int_{\gamma_0} \omega dt'' } e^{ -2i \int_{\gamma_b} \omega dt'' }.
\end{equation}
Assuming we can construct a good steepest descent contour, the $dt'$ integral will be totally dominated for $t'$ near the branch point $t_{\pm}$. Thus the $dt'$ integral will be localized to $\gamma_0$ and will get the exponential's contribution from $\gamma_a$ and $\gamma_0$ only. Moreover, since the time-dependence in $f$ is slow compared to $\omega$ we can set $f(\omega(t'),t') = f(\omega(t'),t_0)$. In other words,
\begin{equation}
I = I_a \int_{\gamma_0} dt' f(\omega(t'),t_0) e^{ -2i \int_{\gamma_0} \omega dt'' }, \ \ \ \ I_a = e^{ -2i \int_{\gamma_a} \omega dt'' }.
\end{equation}
Written as such, we have two integrals to do: $I_a$, which gets us from the asymptotic past up to the singularity, and the $dt'$ integral near the singularity. To define $\gamma$ we need to focus on the latter. What we want is a contour such that the exponential is of the form $e^{u}$ with $u$ real and negative, so that the integral localizes. Also, we want to take the branch cut for $\omega$ such that the exponentials will die off transverse to the cut, i.e. we want the cut along the curve with $| \omega | \approx 0$ near $t_{\pm}$. It turns out that all of this can be achieved by taking the contour straight into the singularity $t_{\pm}$, evading the singularity by a small circle, and then going straight away from the singularity, see the figure. 

Write $t - t_{\pm} = \delta_{\pm} = r_{\pm} e^{i \theta_{\pm}}$, so that we can define the branch of $\omega$ we want by $\omega = \sqrt{r_- r_+} e^{i (\theta_- + \theta_+)/2}$. Near the branch points, the exponential goes like $e^{u_{\pm}}$ where\footnote{All angles are measured from the horizontal as usual.}
\begin{equation}
\label{omega}
u_{\pm} = -2 i \int_{t_{\pm}}^{t} dt \ \omega \approx \Omega_{\pm} \delta_{\pm}^{3/2}, \ \ \ \ \ \Omega_{\pm} = -\frac{4i}{3} \frac{k_*^2}{a_0^2} e^{\pm i \pi/4} \sqrt{| 2 \sigma |}.
\end{equation}
The branch cuts should be taken along local angles $\alpha_{\pm}$ so that the cuts run to $\pm i \infty$, so we have the restrictions
\begin{equation}
\begin{tabular}{ l | c | c }                    
   & (*) & (**) \\
     \hline    
  \text{UHP} & $ 0 < \alpha_+ < \pi$ & $\alpha_+ - 2\pi < \theta_+ < \alpha_+$   \\
    \hline    
  \text{LHP} & $ -\pi < \alpha_- < 0$ & $\alpha_- < \theta_- < \alpha_- + 2\pi$ \\
\end{tabular}
\end{equation}
on the relevant angles. Now, we want to choose $\alpha_{\pm}$ such that $\arg{u_{\pm}}$ is an odd multiple of $\pi/2$ along the cut, and such that we can find two rays emanating from $\delta_{\pm} = 0$ to use as the steepest descent curve (one ray for the incoming and one for the outgoing direction). It turns out this is only possible in the lower half plane. We have $\arg{u_{\pm}} = \arg{\Omega_{\pm}} + \frac{3}{2} \arg{\delta_{\pm}}$, so from the condition (*) we see that in the upper half plane we can only take the cut along $\alpha_+ = \pi/2$, while on the lower half plane we must take the cut along $\alpha_- = -\pi/2$. The steepest descent paths are rays emanating from $\delta_{\pm} = 0$ along the angles $\theta_{\pm}$ such that $\arg{u_{\pm}} = (2n-1)\pi$ for some integers $n$, ensuring that $e^{u_{\pm}} \sim e^{-|u_{\pm}|}$. In the upper half plane this means $\theta_{+} = (8n-3)\pi/6$ and in the lower half plane $\theta_{-} = (8n-1)\pi/6$. However, these angles must still obey condition (**) above, so we find that in the UHP we can only use $n=0$ and thus cannot make a good steepest descent contour. In the LHP, we can use both $n=-1$ and $n=0$, corresponding to the rays along $\theta_- = -\pi/6$ and $+7\pi/6$, respectively. Thus we must take our integration contour in the lower half plane.

The conclusion of this discussion is that we define the branch cuts with $\alpha_{\pm} = \pm \pi/2$ (as shown in the figure), and the steepest descent contour is going around the singular point $t_-$, consisting of the (incoming) ray $\arg{\delta_-} = 7\pi/6$, a small arc around the pole at $\delta_- = 0$, and the (outgoing) ray $\arg{\delta_-} = -\pi/6$. With this setup, the integrals near $t_-$ are easy. We expand the factors of $\omega$ in $f$ around $t=t_-$ and obtain some simple complex line integrals. In practice we need the cases $f \sim \dot{\omega}/\omega = 1/\delta_-$ (for the $\beta$ integral (\ref{betaformal})) and $f \sim \omega^{-n} = Q^{-n} \delta_-^{-n/2}$ for the loop integrals, where
\begin{equation}
Q = e^{-i \pi/4} \left( \frac{k_*}{a_0} \right)^2 \sqrt{|2 \sigma|}.
\end{equation}
Elementary complex analysis yields the following:
\begin{equation}
I_0 = \int_{\gamma_0} dt' \frac{\dot{\omega}}{\omega} e^{u_-} = \frac{2 \pi i}{3}, \ \ \ I_1 = \int_{\gamma_0} dt' \frac{e^{u_-}}{\omega} = Q \sqrt{6}(1-i) \Gamma \left( \frac{4}{3} \right) |\Omega_-|^{-1/3}
\end{equation}
\begin{equation}
I_2 = \int_{\gamma_0} dt' \frac{e^{u_-}}{\omega^2}  = Q^2 \frac{4 \pi i}{3}, \ \ \ I_3 = \int_{\gamma_0} dt' \frac{e^{u_-}}{\omega^3} = Q^3 \sqrt{6}(1-i) \Gamma \left( \frac{2}{3} \right) |\Omega_-|^{+1/3}.
\end{equation}

We also need the integral $I_a$ from $t=-\infty$ to $t=t_-$. As we move horizontally, $\omega$ picks up a real part and the exponential just gets a phase. However, going vertically down to $t_-$, the integral gets an imaginary part. Parametrize the vertical coordinate here by $s$, taken from $0$ to $-\sigma$. In our conventions, along the vertical line we have $\omega = (k_*/a_0)^2 \sqrt{\sigma^2 - s^2}$. Thus
\begin{equation}
I_a = \exp{ \left\{ - 2 i \int_0^{-\sigma} i \ ds \  \omega \right\} } = \exp \left\{ -\frac{\pi}{2} \left[ \left( \frac{k}{k_*} \right)^2 - \left( \frac{3}{2} \frac{a_0 H}{k_*} \right)^2  \right] \right\},
\end{equation}
up to a phase. 

Note that here we have only proven this formula for $k \gg 3 a_0 H/2$. For superhorizon modes, the argument goes over almost identically, except that the branch cut is between the (now real) zeroes of $\omega$ and the contour must be adjusted accordingly. One finds that the formula above holds for arbitrary modes, in agreement with mode-matching computations.

\subsection{Evaluating $\beta$ at late times}

With all of this in place, finding the late-time value of $\beta$ is easy. Starting from (\ref{betaformal}) we consider late times $t \rightarrow \infty$ and take the time integrals along the steepest descent contour $\gamma$ we have just constructed,
\begin{equation}
\beta(k,t \rightarrow \infty) = \int_{\gamma} dt' \frac{\dot{\omega}(k,t')}{2 \omega(k,t')} \exp{ \left\{ -2i \int_{-\infty}^{t'} dt'' \omega(k,t'') \right\}}.
\end{equation}
In the notation of the previous section this is just
\begin{equation}
|\beta| = \frac{1}{2} I_a I_0 = \frac{\pi}{3} \exp \left\{ -\frac{\pi}{2} \left[ \left( \frac{k}{k_*} \right)^2 - \left( \frac{3}{2} \frac{a_0 H}{k_*} \right)^2  \right] \right\}.
\end{equation}
Our argument does not fix the phase of $\beta$; this phase is not relevant for many observables, in particular the ``occupation number'' of the $\chi$ field seen by a late time observer is $n_k = |\beta(k)|^2$.

\subsection{Tadpole diagram (A)}

We can extend the technique described in the previous section to compute the loop diagrams from section 3. First, for each loop momentum $q$, we take the time integral to go from past infinity $t \rightarrow -\infty$ to very late times $t \rightarrow +\infty$ along a contour $\gamma$ in the complex $t$-plane such that $| \dot{\omega}/\omega^2 | \ll 1$ along the entire curve. This ensures that the $\chi$ mode functions can be written in terms of the adiabatic mode functions (\ref{chilate}). Due to this, we can extract the parts of the diagram proportional to $\beta$, i.e. the contribution from $\chi$ excitations. Defining the contour as such, we may deform it continuously to the steepest descent contour $\gamma$ constructed above. After doing the time integrals, we are left with the integrals over loop momenta $q$, which will contain factors of $\beta \sim e^{-q^2/k_*^2}$ and will thus converge without any need for regularization. For both diagrams our arguments are valid only up to a phase, so what we are really going to do is over-estimate all of the contributions by computing the modulus squared of each diagram and adding them coherently.

Let us begin with the tadpole diagram. As there, we label the external graviton's (comoving) momentum by $k$ and the internal $\chi$ propagators by $q$. Starting from eq. (\ref{tadpole}), we take the external time $t \rightarrow \infty$, and deform the time integral to the contour $\gamma$, so the contribution to the correlator is
\begin{equation}
A = - i \int \frac{d^3q}{(2\pi)^3} \int_{\gamma} dt' \ f_4(q,t') g_4(k,t') | \chi_q(t') |^2.
\end{equation}
Here $f_4$ is given by (\ref{f4}), the graviton portion is
\begin{equation}
\label{g4}
g_4(k,t) = \text{Im} \left( h_k(\infty) \overline{h_k(t)} \right)^2 = \frac{H^4}{2 M^4 k^6} [ (1- \frac{\lambda^2}{4})\sin \lambda - \lambda \cos \lambda ], \ \ \ \ \lambda = \frac{2k}{aH},
\end{equation}
and using (\ref{chi0}), (\ref{chilate}) and the Wronskian condition $|\alpha|^2 - |\beta|^2 = 1$ for $\chi$ we have
\begin{equation}
\label{chi2}
| \chi_q(t) |^2 = \frac{1}{2 a^3 \omega} \left[2 \text{Re} \left( \beta \sqrt{1+\beta^2} e^{-2i \int^t dt' \omega(t')} \right)  + 2|\beta|^2 +1 \right].
\end{equation}
Let us consider the scaling in momentum (at high momentum) of each of these terms. We have $f_4 \sim q^2$ (or slower), $\omega \sim q$, and the phase space integral gives $q^2$. Thus, if the momentum integral is cut off at some scale $\Lambda$, the last term in brackets will diverge as $\Lambda^4$ at leading order. However, the other two terms are finite due to $\beta \sim e^{-q^2/k_*^2}$. Moreover, we are trying to study the effect of the presence of excitations, i.e. non-zero $\beta$. Therefore, we will discard the first term from here out and focus on the finite contributions from the latter two terms.

Consider the integral from the first term in (\ref{chi2}),\footnote{We first focus on the terms from $f_4^a$ because the exposition is simpler; the results from $f_4^b$ are identical.}
\begin{equation}
A_{1} = -i \int \frac{d^3q}{(2\pi)^3} \int_{\gamma} dt' \ f_4^a(q,t') g_4(k,t') \beta(q,t') \sqrt{1+\beta(q,t')^2} \frac{e^{-2i \int^{t'} dt'' \ \omega(q,t'')}}{a(t')^3 \omega(q,t')}.
\end{equation}
Here $\beta = \beta(q,t) = 0$ for $t<t_0$ and becomes non-zero at $t_0$. Again assuming that the timescale of non-adiabaticity is much shorter than the Hubble scale, we can take the time arguments of $f_4^a, g_4,$ and $a$ to be $t_0$ since they are all varying much more slowly than $\omega$, and so we just need the time integral $I_1$ from the first section of the appendix. Explicitly (dropping subleading terms in $\beta \ll 1$)
\begin{eqnarray}
|A_{1}| & = & \int \frac{d^3q}{(2\pi)^3} f_4(q,t_0) g_4(q,t_0) \beta(q,t_0) \times \frac{Q I_a I_1}{a_0^3} \\
 & = & \mathcal{O}(1) \times \frac{H^2}{2 M^2 k^3} \frac{k_*^3}{H M^2} F(\lambda_0),
\end{eqnarray}
where $\lambda_0 = 2k/a_0 H$ and $F$ is the dimensionless $\mathcal{O}(10^{-1})$ function
\begin{equation}
\label{F}
F(\lambda_0) = \frac{1}{\lambda_0^{3}} \left[ \left(1-\frac{\lambda_0^2}{4} \right) \sin \lambda_0 - \lambda_0 \cos \lambda_0 \right].
\end{equation}
Note that this function is peaked for $k/a_0 H \sim 1$, i.e. for graviton modes $k$ exiting the horizon at $t=t_0$, see fig. (\ref{FandG}).

Now we need to do the loop integral involving the second term in (\ref{chi2}). This term does not have an oscillating phase but can be easily bound from above. To define the integral in the first place, we do the same trick as before, taking the time integral to run along a large semicircle in the lower half plane so that we can use the adiabatic mode functions. Having done so, consider the integrand
\begin{equation}
A_{2} = -i \int \frac{d^3q}{(2\pi)^3} \int_{\gamma} dt \frac{f_4(t) g_4(t)}{a^3(t) \omega(t)} | \beta(q) |^2.
\end{equation}
This integrand has no branch cuts, but still has the poles where $\omega = 0$. Now, we may deform the contour onto the real time axis, plus a circle around this pole. The contribution from the pole vanishes as we shrink the circle, since near the pole $\omega \sim \delta^{1/2}$ so the integrand scales with the circle's radius as $\sim R^{1/2} \rightarrow 0$. Thus we only have to do the integral on the real axis. Again $\beta$ is zero until $t > t_0$. After $t_0$, the integrand is bounded from above by its value at $t_0$ ($\omega$ is increasing and the other functions are redshifting), and decays to zero within a few multiples of the non-adiabaticity timescale $\Delta t$, so we can overestimate the time integral by taking the value at $t_0$ and multiplying by $\Delta t = k_*^{-1}$. We can then perform the momentum integral. The result is
\begin{equation}
|A_2| \lesssim \mathcal{O}(1) \times \frac{H^2}{2 M^2 k^3} \frac{k_*^3}{H M^2} F(\lambda_0),
\end{equation}
exactly the same as the other term in the diagram, up to a numerical coefficient.

\subsection{Diagram B}

Finally, we need to evaluate the second diagram topology. There are four diagrams here: RR, LL, RL and LR. We will use the same techniques from the previous two sections, although this diagram introduces the additional complication that it has two internal time coordinates $t'$ and $t''$. However, the loop integrand is peaked along the diagonal $t' \approx t''$ due to redshifting and  relative phases. Once we reduce the integrals to ones over only $t'$, we can apply the steepest descent method to localize them to $t' \approx t_0$ as in the previous section.

With this in mind, we start with eq. (\ref{diag2-1}) and (\ref{diag2-2}). Once again taking the external time $t \rightarrow \infty$ and taking both time integrals along a large contour in the complex $t',t''$-planes, we see that the contribution from the RR diagram plus the LL diagram gives
\begin{equation}
B_{RR+LL} = \int \frac{dt' dt'' d^3q}{(2\pi)^3} f_3(q,t',t'') g_3(k,t',t'') h_3(q,q',t',t'')
\end{equation}
where here $f_3$ is given by (\ref{f3}), the momenta satisfy $\mathbf{k} + \mathbf{q} + \mathbf{q'} = 0$, $g_3$ comes from the graviton propagators
\begin{equation}
g_3(k,t',t'') = \text{Re} \left( h_k(\infty)^2 \overline{h_k(t')} \overline{h_k(t'')} \right)
\end{equation}
and $h_3$ comes from the $\chi$ propagators,
\begin{equation}
\label{h3}
h_3(q,q',t',t'') = \chi_q(t') \overline{\chi_q(t'')} \chi_{q'}(t') \overline{\chi_{q'}(t'')}.
\end{equation}
The momentum integral is going to be dominated by the internal momenta $q \approx k_*$ while the external momentum is only relevant around $k \approx a_0 H \ll k_*$. Thus we may approximate $q' \approx q$. With this approximation, let us insert the adiabatic mode functions (\ref{chi0}), (\ref{chilate}) into (\ref{h3}). We obtain
\begin{eqnarray*}
\lefteqn{ h_3(q,t',t'') = \frac{1}{4 a^3(t') a^3(t'') \omega(t') \omega(t'')} \Big[ 4 |\alpha|^2 |\beta|^2 } \\
& & \mbox{} + 2 |\alpha|^2 \alpha \beta e^{-2i\theta'} + 2 \overline{\alpha} \beta |\beta|^2 e^{+2i\theta'} + 2 \alpha \beta |\beta|^2 e^{-2i\theta''} + 2|\alpha|^2 \overline{\alpha} \overline{\beta} e^{+2i\theta''} \\
& & \mbox{} + |\alpha|^4 e^{-2 i \Delta \theta} + |\beta|^4 e^{+2 i \Delta \theta} \\
& & \mbox{} + \alpha^2 \beta^2 e^{-2 i \theta'} e^{-2 i \Delta \theta} + \overline{\alpha}^2 \overline{\beta}^2 e^{+2i\theta'} e^{+2i \Delta \theta} \Big].
\end{eqnarray*}
Here the phases are
\begin{equation}
\theta' = \int_{-\infty}^{t'} d\tau \ \omega(\tau), \ \ \ \theta'' = \int_{-\infty}^{t''} d\tau \ \omega(\tau),  \ \ \ \Delta \theta = \theta' - \theta'' = \int_{t'}^{t''} d\tau \ \omega(\tau).
\end{equation}

This somewhat fearsome expression for $h_3$ shows that there are four types of terms, organized by line. The first term has no oscillating phase but is straightforward to estimate using the same logic as $A_2$ above. It gives a contribution
\begin{equation}
B_{RR+LL}^{1} \approx \int \frac{dt' dt'' d^3q}{(2\pi)^3} \frac{f_3(q,t',t'') g_3(k,t',t'')}{a^3(t') a^3(t'') \omega(t') \omega(t'')} |\alpha|^2 |\beta|^2.
\end{equation}
As explained earlier, the $t''$ integral will localize to $t'' \approx t'$, so we may set $t'' = t', \int dt'' \rightarrow \Delta t$. The remaining integral over $t'$ has zero integrand for $t' < t_0$ and the integrand is bound from above by its value at $t' = t_0$ (again because the frequency increases and the rest of the functions are redshifting). Thus we can set all the time arguments to $t_0$ and take $\int dt' \rightarrow \Delta t$, obtaining an easy momentum integral which yields
\begin{equation}
| B_{RR+LL}^{1} | \lesssim \mathcal{O}(1) \times \frac{H^2}{2 M^2 k^3} \frac{k_*^3}{H M^2} G_{RR+LL}(\lambda_0)
\end{equation}
which has the same parametric dependence as the previous diagram, but a different dimensionless function from the graviton propagators,
\begin{equation}
G_{RR+LL}(\lambda_0) = \frac{1}{\lambda_0^3} \left[ \left(1 - \frac{\lambda_0^2}{4} \right) \cos \lambda_0 + \lambda_0 \sin \lambda_0 \right].
\end{equation}
Here once again $\lambda_0 = 2k / a_0 H$. Although this function is divergent as $\lambda_0 \rightarrow 0$, we will see that the total contribution from all four diagrams gives a finite result.

The second type of term goes over similarly to the previous diagram. Consider for example the first such term, which has the phase $\sim \exp(-2i \int \omega)$, just like the oscillating terms from the previous diagram. As argued earlier, the $t''$ integral is dominated near $t'' \approx t'$ so again we set $t'' = t', \int dt'' \rightarrow \Delta t$. Other than the phase $\theta' = \int^{t'} \omega$ all other time-dependent functions are varying slowly, and we can localize the $t'$ integral to a steepest descent contour near $t' = t_-$ as before, so we get (again dropping subleading terms in $\beta$)
\begin{equation}
B_{RR+LL}^{2} \approx g_3(k,t_0) a_0^6 \Delta t \int \frac{dt' d^3q}{(2\pi)^3} \beta \frac{e^{-2i \int^{t'} \omega}}{\omega^2(q,t)}.
\end{equation}
This time integral is $I_2$ from above. Using that result we are left with an easy momentum integral that yields\begin{equation}
|B_{RR+LL}^{2}| \lesssim \mathcal{O}(1) \times \frac{H^2}{2 M^2 k^3} \frac{k_*^3}{H M^2} G_{RR+LL}(\lambda_0),
\end{equation}
just like the first term.

Terms from the third line require a bit of extra work, although they actually give the same result. We need to be a little delicate with the $t''$ integral. This should still be dominated by $t'' \approx t'$, but now we note that in this approximation
\begin{equation}
\int^{t'}_{-\infty} dt'' e^{-2 i \Delta \theta} = \int^{t'}_{-\infty} dt'' e^{-2 i \int_{t'}^{t''} d\tau \ \omega(\tau)} \approx \frac{i}{2 \omega(t')}.
\end{equation}
After estimating the $t''$ integrals in this fashion, we have reduced everything to the case from the previous paragraph and again we obtain answers of the form
\begin{equation}
|B_{RR+LL}^{3}| \lesssim \mathcal{O}(1) \times \frac{H^2}{2 M^2 k^3} \frac{k_*^3}{H M^2} G_{RR+LL}(\lambda_0).
\end{equation} 
Finally, to evaluate the terms in the last line, we essentially combine the arguments from the previous two types of term. Namely, the $t''$ integral is replaced by $\int dt'' \rightarrow i/2 \omega(t')$, leaving us with a time integral of the form
\begin{equation}
\int dt' \frac{e^{-2 i \int^{t'} \omega}}{\omega^3(t')} = I_3.
\end{equation}
Using this and the usual arguments about $t' \approx t_0$ we again obtain momentum integrals which evaluate to
\begin{equation}
|B_{RR+LL}^{4}| \lesssim \mathcal{O}(1) \times \frac{H^2}{2 M^2 k^3} \frac{k_*^3}{H M^2} G_{RR+LL}(\lambda_0).
\end{equation}

In summary, what we have found so far is that all the terms from the right-right and left-left diagrams give the same type of contribution, 
\begin{equation}
|B_{RR+LL}| \lesssim \mathcal{O}(1) \times \frac{H^2}{2 M^2 k^3} \frac{k_*^3}{H M^2} G_{RR+LL}(\lambda_0).
\end{equation}
Now, we need to do the right-left and left-right diagrams. These are in fact identical so we only need to do one. Writing out the integrand starting with (\ref{diag2-2}) and following the same steps used to simplify it as in the previous case, one sees immediately that the integral is exactly the same as the RR case, except that the graviton propagators are different
\begin{equation}
g_3 \rightarrow \tilde{g}_3 = |h_k(\infty)|^2 h_k(t') \overline{h_k(t'')}.
\end{equation}
The effect of this is that the combined diagrams give
\begin{equation}
|B_{RL+LR}| \lesssim \mathcal{O}(1) \times \frac{H^2}{2 M^2 k^3} \frac{k_*^3}{H M^2} G_{RL+LR}(\lambda_0),
\end{equation}
where
\begin{equation}
G_{RL+LR}(\lambda_0) = \frac{1}{\lambda_0^3} \left(1 + \frac{\lambda_0^2}{4} \right)
\end{equation}
and the coefficient we have dropped is the same as the $RR+LL$ case. The relative negative sign between the $RR$ and $RL$ diagrams (\ref{diag2-1}), (\ref{diag2-2}) then conspires to make the total result finite at $\lambda_0 \rightarrow 0$, and we find that the total contribution to the tensor two-point function from the $B$ diagrams is
\begin{equation}
|B| \lesssim \mathcal{O}(1) \times \frac{H^2}{2 M^2 k^3} \frac{k_*^3}{H M^2} G(\lambda_0),
\end{equation}
where again $\lambda_0 = 2k/a_0 H$ and $G$ is
\begin{equation}
\label{G}
G(\lambda_0) = \frac{1}{\lambda_0^3} \left[ 1 + \frac{\lambda_0^2}{4} - \left( 1 - \frac{\lambda_0^2}{4} \right) \cos \lambda_0 - \lambda_0 \sin \lambda_0 \right].
\end{equation}
This function is $\mathcal{O}(10^{-1})$ and peaked for $k \approx a_0 H/4$, similar to the function $F(\lambda_0)$ appearing in the other diagram; see figure (\ref{FandG}).

\bibliographystyle{JHEP} 
\bibliography{draft-shorter}

\end{document}